\documentclass[aps]{revtex4}

\usepackage{graphicx}
\usepackage{subfigure}
\usepackage{epsfig}
\usepackage{dcolumn}
\usepackage{bm}

\begin{document}

\newcommand{\pr}{\partial}
\newcommand{\ep}{\epsilon}
\newcommand{\p}{\prime}
\newcommand{\w}{\omega}
\newcommand{\lt}{\mathcal{L}}
\newcommand{\Fa}{F\left[\begin{array}{c|c} x & x^\p \\ t+\Delta t & t \end{array} \right]}
\newcommand{\Fb}{F\left[\begin{array}{c|c} x_t & x_0 \\ t & t_0 \end{array} \right]}

\newcommand{\nwc}{\newcommand}
\nwc{\beq}{\begin{equation}}
\nwc{\eeq}{\end{equation}}
\nwc{\bdm}{\begin{displaymath}}
\nwc{\edm}{\end{displaymath}}
\nwc{\bea}{\begin{eqnarray}}
\nwc{\eea}{\end{eqnarray}}
\nwc{\para}{\paragraph}
\nwc{\vs}{\vspace}
\nwc{\hs}{\hspace}
\nwc{\la}{\langle}
\nwc{\ra}{\rangle}
\nwc{\pref}{\pageref}
\nwc{\lw}{\linewidth}
\nwc{\nn}{\nonumber}

\large

\title{Energy fluctuations in a biharmonically driven nonlinear system}

\author{Navinder Singh, Sourabh Lahiri and A. M. Jayannavar}

\email{jayan@iopb.res.in}

\affiliation{\large Institute of Physics, Sachivalaya Marg, Bhubaneswar-751005, India}

\begin{abstract}
We study the fluctuations of work done and dissipated heat of a Brownian particle in a symmetric double well system. The system is driven by two periodic input signals that rock the potential simultaneously. Confinement in one preferred well can be achieved by modulating the relative phase between the drives. We show that in the presence of pumping the stochastic resonance signal is enhanced when analyzed in terms of the average work done on the system per cycle. This is in contrast to the case when pumping is achieved by applying an external static bias, which degrades resonance. We analyze the nature of work and heat fluctuations and show that the steady state fluctuation theorem holds in this system.
\end{abstract}                                           

\maketitle


	
\section{Introduction}

A number of important results have been obtained over the last two decades on the statistical properties of fluctuations in physical quantities in non-equilibrium processes. These are referred to as Fluctuation Theorems (FTs) \cite{Bustamante,evans02,harris07,ritort,kur07,dje93,dje94,galco95,kur98,lebo99,crooks99,crooks00,udo,jar,zon02,zon04,abhi}. They allow extension of thermodynamic concepts to small systems \cite{Bustamante}. The fluctuation theorems reveal rigorous relations for the properties of distribution functions of physical variables such as work done, dissipated heat or entropy production for systems driven far away from equilibrium, independent of the nature of driving. They are not restricted to the linear response regime, thus allowing us to obtain results generalizing Onsager reciprocity relations to the nonlinear response coefficients in nonequilibrium state. From these theorems, corollaries such as the statistical derivation of the second law, can be established. There are different fluctuation theorems, depending on the physical quantities they relate to and on the state of the system they refer to.  These theorems are useful to probe nonequilibrium states in nanophysics and biology.
Hence they are anticipated to play an important role in the design of nanodevices and engines (molecular motors) at nanoscales. The distributions of heat and work in relation to FTs have been experimentally studied for few Brownian systems
 \cite{cil98,wang02,lip,trep,dou,joub}. 

\vspace{0.2cm}

In recent theoretical \cite{saik,mam} and experimental \cite{cili} studies, the distributions of dissipated heat and work done on the system have been explored in a system exhibiting stochastic resonance \cite{saik,mam}. The steady state fluctuation theorem (SSFT) (as elaborated upon in section \ref{sec:3F}) holds in this system. Exploring the FTs in nonlinear systems by changing the symmetry of the driving force cycle has been suggested in Ref. \cite{cili}. To this end, we study the dynamics of a particle in a symmetric double well potential which is in contact with a thermal bath at temperature $T$. This system  exhibits stochastic resonance (SR) under subthreshold external ac drive \cite{parisi}. The fundamental periodic component of the system response (i.e., the amplitude of the feeble input at the same frequency) can be amplified  by the assistance of noise. It is reflected as a peak in the output signal-to-noise ratio as a function of noise strength. This peak occurs when the noise induced switching rate in the system matches the forcing frequency. This optimization or synchronization condition is achieved by tuning the noise intensity. This phenomenon is known as {\emph{ stochastic resonance}} \cite{parisi,gam}. Here noise plays a constructive role as opposed to our conventional wisdom that the presence of noise degrades the signal. This is due to a cooperative interplay between the system nonlinearity and input signal in addition to the noise. Because of its generic nature, this phenomenon boasts applications in almost all areas of natural science \cite{gam}. To characterize this resonance phenomenon, several different quantifiers have been introduced in the literature \cite{gam,gamma,evs,mpl,maha97,iwai,mahato,dan}. One of the quantifiers, namely the input energy of the system or the work done on the system per cycle is known to characterize SR as a bona fide resonance \cite{dan,iwai,saik,mam}. In this case, the resonance  can be shown to occur both as a function of noise strength and driving frequency.

It is known that static asymmetry in the bistable potential weakens the magnitude of the SR effect \cite{mam,gam}. Static tilt in the potential makes one potential well more stable than the other leading to more particle localization or pumping in one well (lower well) compared to the other. Moreover, due to asymmetry in the potential, escape rate of a particle from higher to lower well will be different from lower to higher well. These two different rates make synchronization difficult between the signal and the dynamics of the particle hopping, since the driving frequency cannot match both these hopping rates simultaneously.

In the present work, we study the SR for a particle in a symmetric double well potential, driven simultaneously by two periodic signals of frequencies $\omega$ and $2\omega$ with a relative phase difference $\phi $ between them. Such a force  averaged over a period does not lead to a net bias and yet particle is preferentially pumped  into one well depending on  phase difference  $\phi$ and other physical parameters \cite{borro04,savelev,borro05,borro-acta,pla,chi,ajdari}. This phenomenon is  known as \emph{harmonic mixing} \cite{borro04,savelev,borro05,borro-acta}.
Due to this statistical confinement of the particle, similar to the case of static tilt \cite{gam,mam09}, we expect to observe a reduced SR signal in this system. However, contrary to this expectation, we show that the resonance signal is enhanced in the presence of the biharmonic drive at frequency $2\omega$ when analyzed in terms of the input energy (or the work done) as a quantifier of SR. Using stochastic energetics \cite{seki,udo,parrando} we also study the  nature of fluctuations in the work done,  dissipated heat and internal energy across SR. In some range of parameters, nature of hysteresis loops is analyzed. We show that the SSFT holds for work done over a long time interval. The modified SSFT for heat is also studied in this system.

\section{The Model: Brownian particle in a Rocked Double Well Potential}
We consider the stochastic dynamics of a Brownian particle in a double-well potential $V(x) = -\frac{ x^2}{2}+\frac{ x^4}{4}$, rocked by a weak biharmonic (time-asymmetric) external field $F(t) = A \cos(\omega t)+B \cos(2 w t + \phi)$. The potential $V(x)$ has two minima at $x = \pm 1$, separated by a central (at $x = 0$) potential barrier of height $\Delta V = 0.25$. The overdamped Langevin dynamics is given by \cite{risk},

\beq
\gamma \frac{dx}{dt} = -\frac{\partial U(x,t)}{\partial x} + \xi(t),
\label{Lang1}
\eeq

where $U(x,t) = V(x) - xF(t)$, $\gamma$ is the friction coefficient, $\xi(t)$ is the Gaussian white noise with the properties
\bea
\la\xi(t)\ra &=& 0 ,\nn\\
\la\xi(t)\xi(t')\ra &=& 2 D \delta(t-t') ,
\eea
where $D= \gamma k_BT$. 
The thermodynamic work done by an external drive over a period  $\tau_\omega (=\frac{2\pi}{\omega})$ is given by \cite{seki,udo,parrando}

\bea
W_p &=& \int_{t_0}^{t_0+\tau_\omega} \frac{\partial U(x,t)}{\partial t} dt \nn\\
&=& \int_{t_0}^{t_0+\tau_\omega} x(t) [A\omega\sin \omega t + 2B\omega\sin(2\omega t + \phi)]dt.
\eea

This work (or input energy) over a period equals the change in the internal energy $\Delta U_p = U(x(t_0+\tau_\omega), t_0+\tau_\omega) - U(x(t_0),t_0)$ plus the heat dissipated over a period $Q_p$, i.e.,

\beq
W_p = \Delta U_p + Q_p.
\label{firstlaw}
\eeq

The above equation is the statement of the First law of thermodynamics and can readily be obtained using  stochastic energetics \cite{seki}. Since $x(t)$ is a stochastic variable, it follows from eq. (3) and eq.(4) that $W_p,\,\,\Delta U_p\,\,$ and $Q_p$ are random variables when evaluated over different periods and realizations of $x(t)$. The above model is solved numerically by using Heun's method \cite{mannela} (all the physical quantities are in dimensionless units). We have ignored the initial transient regime up to time  $t_0$ and evaluated $W_p,\,\, Q_p,$ and $\Delta U_p$ over many cycles ($\sim 10^5$)  of a single long trajectory of the particle.

\section{Results and Discussions}

\subsection{SR as a function of noise strength}
\begin{figure}
\centering
\epsfig{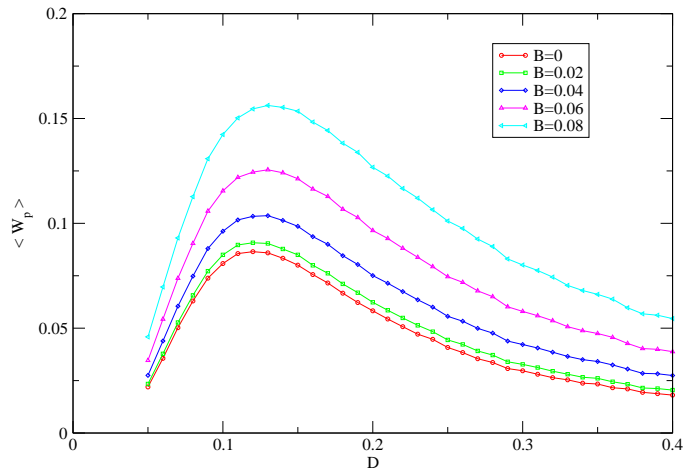} 
\caption{The input energy $\la W_p\ra$ as a function of $D$ for different values of the strength of second harmonic $(B)$. The parameters are: $\omega = 0.1, A = 0.1$, and $\phi = 0$.}
\label{WD}
\end{figure}

\begin{figure}
\centering
\epsfig{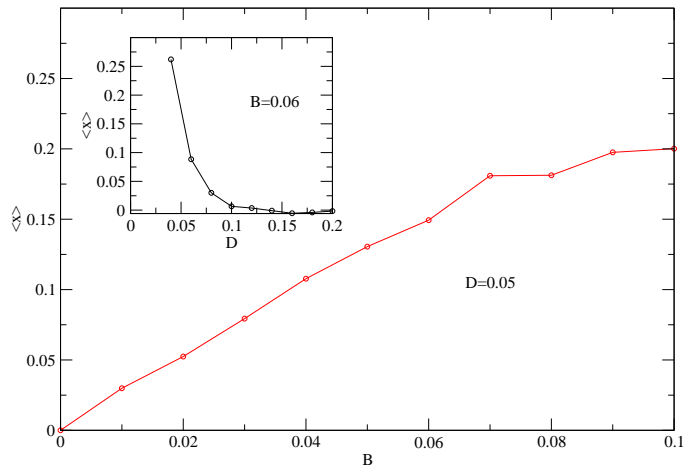} 
\caption{Particle mean position $\la x\ra$ as a function of $B$ for $D=0.05$.  In the inset we have plotted $\la x\ra$  as a function  $D$. Other parameters are: $B=0.06, A=0.1,$ and $\omega=0.1$.}
\label{xbarD}
\end{figure}

In figure \ref{WD}, we have plotted the average work done over a single period $\la W_p\ra$ in the time asymptotic regime as a function of noise strength $D$ for different values of biharmonic drive strength $B$ (for A=0.1). Phase difference is taken to be zero. Other parameters are mentioned in the figure captions. For the case of $B=0$ we have reproduced earlier results \cite{iwai,dan,mam,saik}.
The average input energy ($\la W_p\ra$) shows a peak signifying stochastic resonance (SR) as discussed extensively in earlier literature \cite{iwai,dan,mam,saik}. The quantity $\la W_p\ra$ can also be identified as the average dissipated heat or hysteresis loss into the bath in a time periodic steady state. This follows from eq. (4) by noting that the internal energy being a state variable, $\Delta U_p$ averaged over a period is identically equal to zero.
As we increase $B$, $\la W_p\ra$ increases for fixed value of $D$. Moreover, for different values of $B$, the system exhibits SR as a function of noise strength. The system in a steady state absorbs energy from the external drive and the same is dissipated as heat into the surrounding medium. Hence it is expected that at the resonance the system will absorb maximum energy from the external drive. The input energy curves for higher values of $B$ lie above those for the lower values of $B$. The graph for $B=0$ matches with the earlier known results \cite{iwai,dan,mam}, and the peak position shifts towards higher values of $D$.
It is evident from the figure that in the presence of biharmonic drive  enhancement of SR signal occurs even though there is more statistical confinement of the particle (as $B$ increases) in one well as shown in figure 2.  In this figure we have plotted average position ($\la x\ra$) over period in the time asymptotic regime as a function of $B$ for fixed $D = 0.05$. The value of $\la x\ra$ not being zero signifies selective pumping or localization of particle from one well to another in the presence of biharmonic drive.  Correspondingly, the probability density distribution of the particle averaged over a period shows a marked asymmetry even  though the potential $V(x)$ is symmetric \cite{borro05}. In the absence of second harmonic component i.e., $B= 0$, $\la x\ra = 0$ as expected. The pumping is very significant at low values of temperature. As we increase temperature, the effective pumping reduces. Around and beyond SR, pumping is quite small as shown in the inset of figure \ref{xbarD}.

Stochastic resonance being a synchronization phenomenon \cite{maha97,mahato} it is expected that particle hopping dynamics between the wells get synchronized with the input signal. We find that the  relative variance  (RV)  in physical quantities such as work $\left[ = \frac{\sqrt{\langle W_{p}^2\rangle - \langle W_{p}\rangle^2}}{\langle W_{p}\rangle}\right]$ and heat $\left[ = \frac{\sqrt{\langle Q_{p}^2\rangle - \langle Q_{p}\rangle^2}}{\langle Q_{p}\rangle}\right]$ also show minima at SR \cite{saik,mam,cili}. 

\begin{figure}
\centering
\epsfig{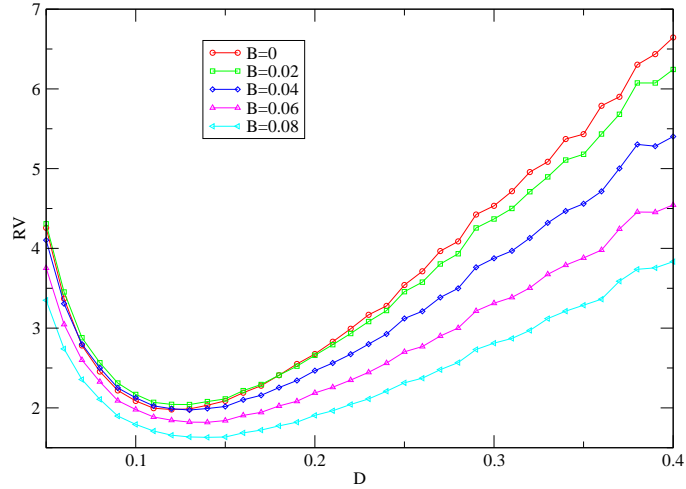} 
\caption{Relative variance (RV) of input energy versus  $D$ for different values of $B$. Other parameters are same as in figure 1.}
\label{rv}
\end{figure}
\begin{figure}
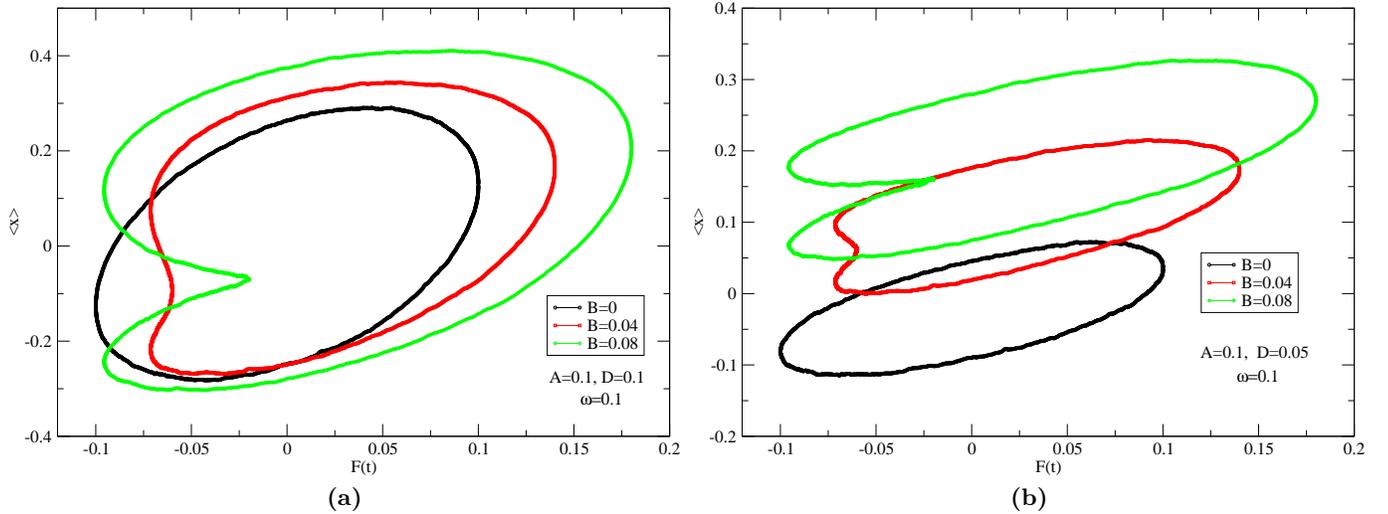

\centering
\begin{tabular}{cc}
\epsfig{file=fig4a.eps,angle=0,width=0.5\linewidth,clip=} &
\epsfig{file=fig4b.eps,angle=0,width=0.5\linewidth,clip=} \\
{\bf (a)} & {\bf (b)}
\end{tabular}
\caption{(a) Hysteresis loops $(\langle x\rangle)~versus~F$ for different values of $B$, and for $D=0.1$. Other parameters are same in figure 1.
(b) Hysteresis loops for different values of $B$ at $D=0.05$.}
\label{hystB}
\end{figure}

In figure \ref{rv} we have plotted relative variance (RV)  as a function of $D$ for various values of $B$. The parameters used are the same as in figure 1. For a given value of $B$ the RV shows a minimum around the same value of $D$ at which $\la W_p\ra$ exhibits a maximum. As the amplitude of the biharmonic drive $B$ increases, RV curves shift downwards. Higher the value of $B$, the lower is the value of RV at the resonance. These results are consistent with figure 1. In the parameter regime that we have considered, the RV is larger than one, i.e., variance in work is large compared to the mean. Hence in this regime, one should analyze full probability distribution as opposed to moments to get better understanding of the phenomenon.

Increasing the amplitude of biharmonic drive  leads to more statistical confinement of particles(figure \ref{xbarD}). This  must be reflected in the nature of hysteresis loops \cite{mahato,pla}. More the pumping, more is the asymmetry in the hysteresis loops, as can be  seen in figure  \ref{hystB}(a) and (b). In these figures hysteresis loops are plotted for different values of $D$ and $B$. The pumping of the particles also gets reflected in the shifting of figures in the vertical upward direction(as $\la x\ra > 0$).
For the case when $B=0$, there will be no pumping and as expected loop is symmetric.
\subsection{SR in the presence of static tilt}
\begin{figure}
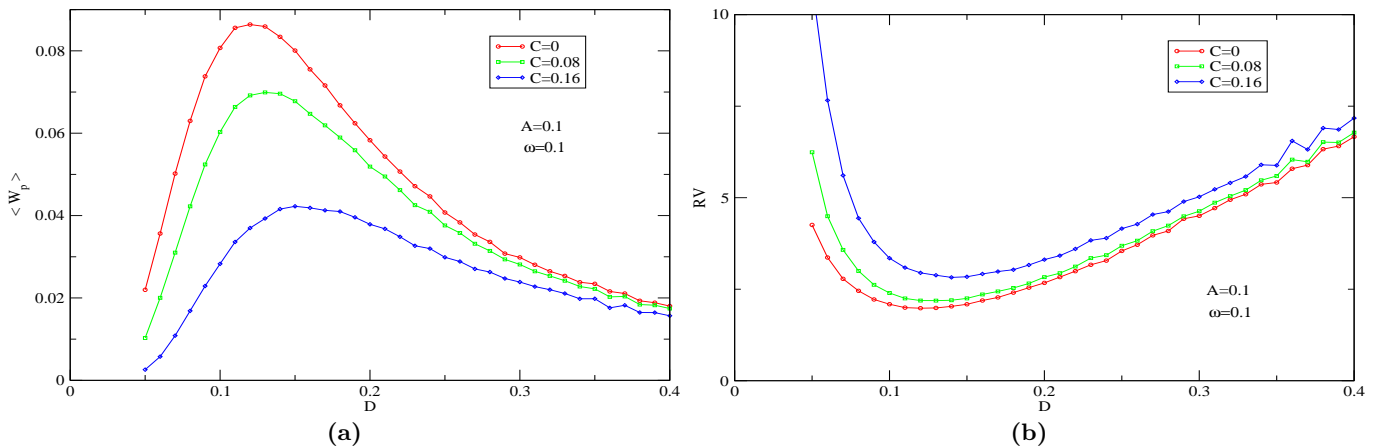

\centering
\begin{tabular}{cc}
\epsfig{file=fig5a.eps,width=0.5\linewidth,height=0.3\lw,clip=} &
\epsfig{file=fig5b.eps,angle=0,width=0.5\linewidth,height=0.3\lw,clip=} \\
{\bf (a)} & {\bf (b)}
\end{tabular}
\caption{(a) Plots of $\la W_p\ra$ as a function of $D$ for different values of the static tilt ($c$). (b) Corresponding plots of relative variance of $W_p$ as a function of $D$. Fixed parameters are mentioned on the graphs. }
\label{WDC}
\end{figure}

\begin{figure}
\vspace{1cm}
\centering
\epsfig{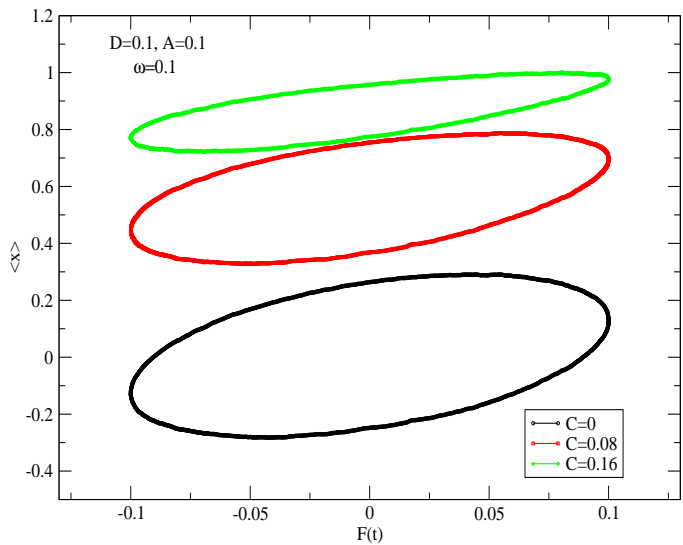} 
\caption{Plots showing hysteresis loops for different values of static tilt $(c)$. Fixed parameters are: $D=0.1, A=0.1$ and $\omega=0.1$. }
\label{xFC}
\end{figure}

Particle pumping in a preferential well can also be induced by applying a static tilt to the otherwise symmetric double well potential. For this we take potential to be $V_1(x) = -\frac{x^2}{2}+\frac{x^4}{4}-cx$. Depending on the value of $c$, the potential $V_1(x)$ becomes asymmetric and obviously more pumping results in the lower potential well. When this system is driven by external AC force $A\cos \omega t$ we show that SR signal weakens.
Figure \ref{WDC}(a) shows the average input energy as a function of $D$ for various values of  $c$. From this, we notice that the input energy curves for higher value of $c$ are below those with lower value of $c$ (other parameters being fixed). As $c$ increases SR peak becomes broadened and shifts towards higher values of $D$. We thus observe that in the presence of pumping induced by static tilt, SR  weakens as mentioned in the introduction. This is also corroborated by the nature of relative work fluctuations  as a function of $D$ (figure \ref{WDC}(b)). From this figure we note that as we increase c RV increases for a given value of $D$. The magnitude of the RV at the minimum becomes larger as we increase $c$. This implies degradation of SR signal in the presence of particle pumping induced by a static tilt. The pumping due to static tilt makes the hysteresis loops asymmetric(figure \ref{xFC}). By increasing c more pumping is achieved and this is reflected in the vertical shift of hysteresis loops. Thus from the above figures and discussions, we conclude that in the presence of biharmonic drive, SR increases while in the presence of static tilt, SR weakens.
\subsection{SR as a function of driving frequency}
\begin{figure}
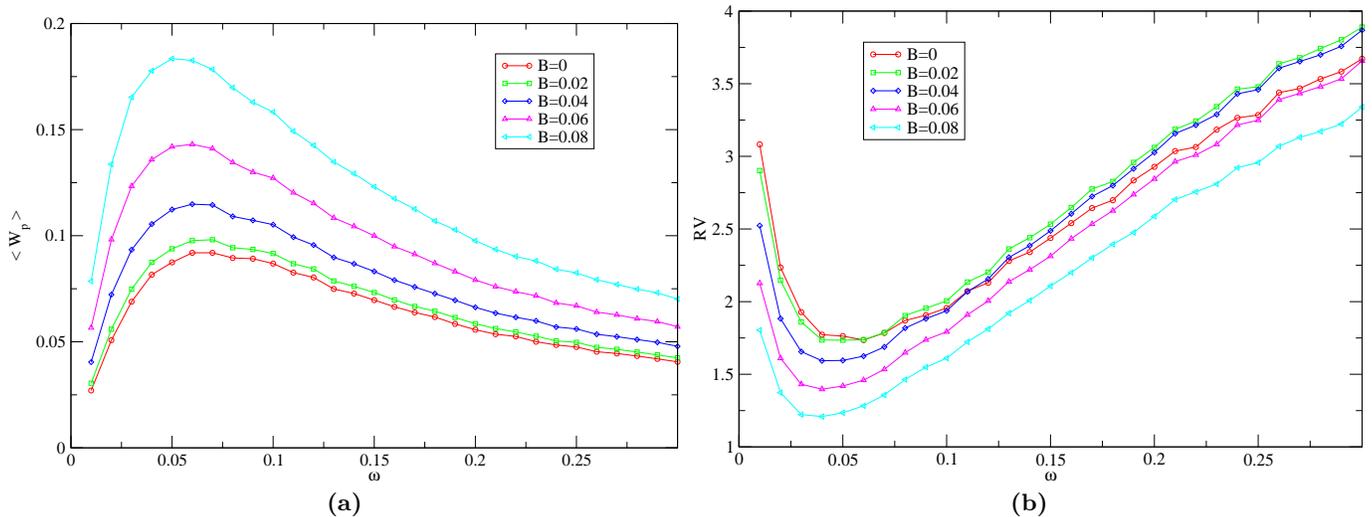

\centering
\begin{tabular}{cc}
\epsfig{file=fig7a.eps,width=0.5\linewidth,clip=} &
\epsfig{file=fig7b.eps,width=0.5\linewidth,clip=}\\
{\bf (a)} & {\bf (b)}
\end{tabular}
\caption{(a) The average input energy per period $\la W_p\ra$as a function of frequency $\omega$. 
(b) Relative variance (RV) of input energy vs frequency $\omega$ for different values of the strength of the second harmonic, $B$. The parameters used are: $\omega = 0.1, A = 0.1, \phi = 0$.}
\label{WomB}
\end{figure}
In figure \ref{WomB}(a), we have plotted average input energy as a function of $\omega$ for various values of $B$. Once again we notice that SR signal even for this case is increased as we increase the biharmonic component $B$. Each curve exhibits a peak as a function of $\omega$, thus establishing SR as a bona fide resonance. The peak shifts to the lower values of $\omega$ as we increase $B$. This is consistent with the fact that peaks in figure \ref{WD} shift towards larger values of $D$ as we increase $B$. This is a requirement for the time scale matching between $D$ and $\omega$. Since increase in $B$ slows down the effective time averaged hopping rates between the wells, hence higher $D$ is required to achieve resonance. This lowering of effective escape rate at given $D$ in turn implies decrease in the resonant frequency. The enhancement of SR signal in the presence of B can be inferred from figure \ref{WomB}(b) where we have plotted relative variance across the SR as a function of $\omega$ for various values of $B$. Lower values of relative variance across the SR for larger values of $B$ are suggestive of the fact that SR is enhanced in the presence of biharmonic drive, consistent with the conclusions of figure \ref {WD}.
\subsection{Energy fluctuations over a single period}
\begin{figure}
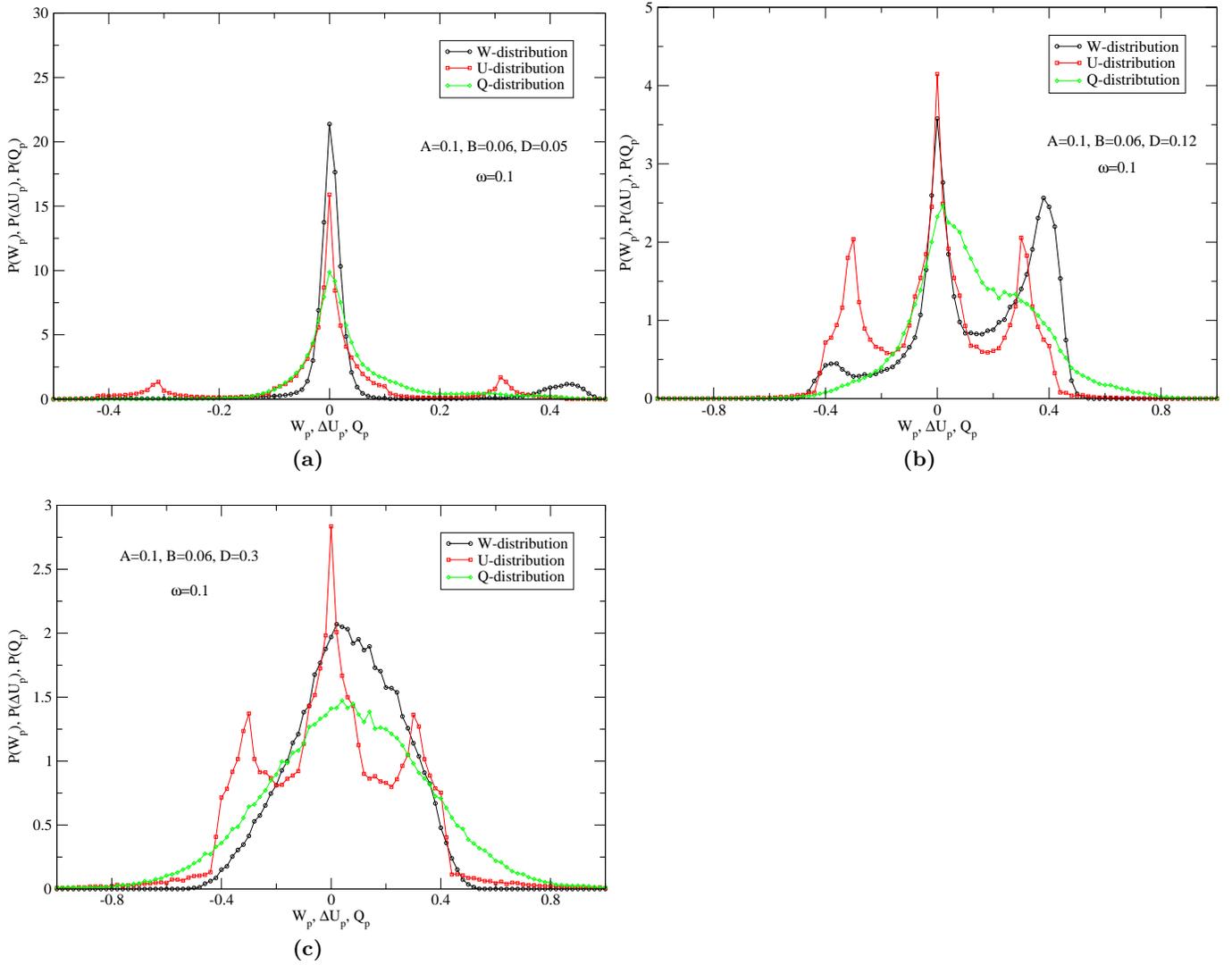

\centering
\begin{tabular}{cc}
\epsfig{file=fig8a.eps,angle=0,width=0.5\linewidth,clip=}&
\epsfig{file=fig8b.eps,angle=0,width=0.5\linewidth,clip=}\\
{\bf (a)} & {\bf (b)} \\\\
\epsfig{file=fig8c.eps,angle=0,width=0.5\linewidth,clip=}\\
{\bf (c)}
\end{tabular}
\caption{Plots (a), (b), and (c) show the distributions $P(W_p)$, $P(Q_p)$, and $P(\Delta U_p)$ for different values of $D$, below resonance $D=0.05$, at resonance $D=0.12$, and above resonance $D=0.3$, respectively. Other fixed parameters are also shown on the graphs.}
\label{fig:WUQ}
\end{figure}
Next, we analyze the nature of distribution functions of input energy $P(W_p)$, dissipated heat $P(Q_p)$ and internal energy $P(\Delta U_p)$ for different values of $D$.  These distributions are plotted in figure \ref{fig:WUQ} (a), (b), and (c) below resonance ($D=0.05$), at resonance ($D=0.12$), and above resonance ($D=0.3$) respectively. The averaged internal energy $\la U\ra$ being a state function assumes the same value at the beginning and at the end of a period or periods in the time asymptotic regime. Hence total change in the internal energy $\la \Delta U_p\ra$ average over a period is equal to zero and it is also expected that the distribution $P(\Delta U_p)$ is symmetric as is evident from figure \ref{fig:WUQ}(a),(b), and (c). The nature of $P(\Delta U_p)$ is explained in  \cite{cili} for a single harmonic drive.
As opposed to $\Delta U_p$, distributions for $W_p$ and $Q_p$ are asymmetric. These distributions keep on changing in shape depending on the number of cycles over which they have been obtained which will be discussed later in connection with steady state fluctuation theorem (SSFT). Probability distributions for work and heat have finite weights for the negative values of their arguments. These negative values correspond to the trajectories where the particle moves against the perturbing AC field over a short time. For small values of $D$($D=0.05$), peak for $W_p$ or $Q_p$ near the origin corresponds mainly to intrawell dynamics of the particle and is mostly confined to a single well.  The occasional excursion of the particle into the other well as a function of time is clearly reflected as a small hump at higher values of $W_p$ or $Q_p$ in the plot of $P(W_p)$ and $P(Q_p)$. As we increase $D$ further ($D=0.12$), interwell dynamics starts playing a dominant role, and hence the distributions become broader. Work distribution exhibits three prominent peaks including one at the negative side. For larger values of $D$ beyond SR point, shapes of $P(W_p)$ and $P(Q_p$) tend closer to Gaussian distribution with increased variance/fluctuations. For such high temperatures, particle makes several random excursions between the two wells during a single time period of the external drive. It may be noted that the relative variances in $W_p$ and $Q_p$ are larger than 1. Also, fluctuations in heat are larger than those of work when averaged over a single period.
\subsection{Effect of phase difference on SR}
\begin{figure}
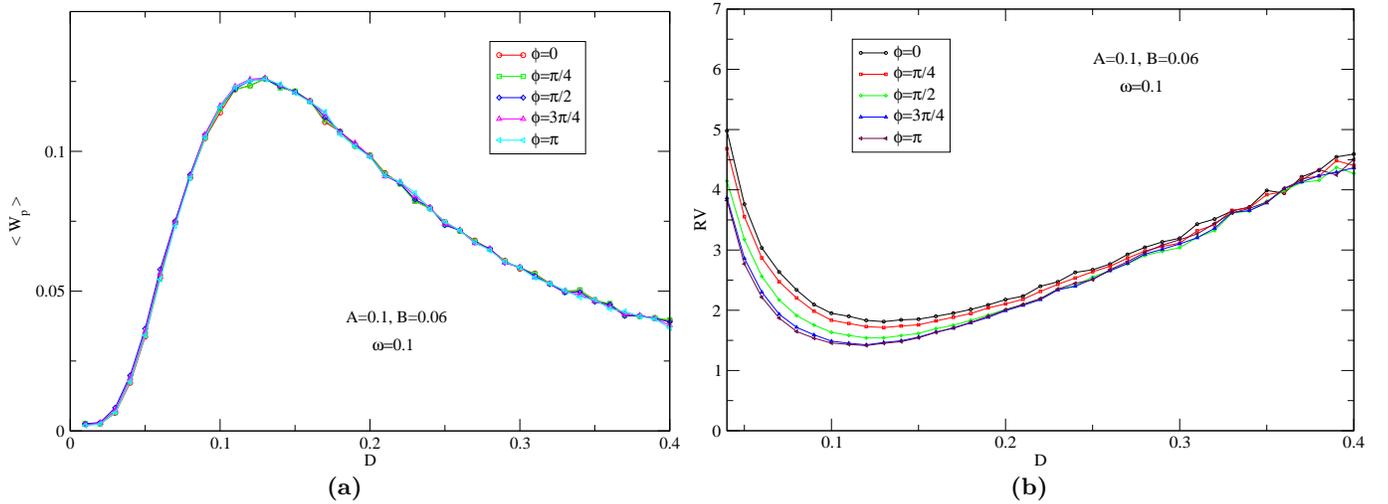

\centering
\begin{tabular}{cc}
\epsfig{file=fig9a.eps,width=0.5\linewidth,clip=} &
\epsfig{file=fig9b.eps,width=0.5\linewidth,height=0.345\lw,clip=} \\
{\bf (a)} & {\bf (b)}
\end{tabular}
\caption{(a) The average input energy per period $\la W_p\ra$as a function of $D$ and frequency $\omega$ for various values of the phase difference $\phi$. (b) relative variances (RV) of $W_p$ versus $D$. Fixed parameters are shown on the graphs.}
\label{fig:WDphi}
\end{figure}
\begin{figure}
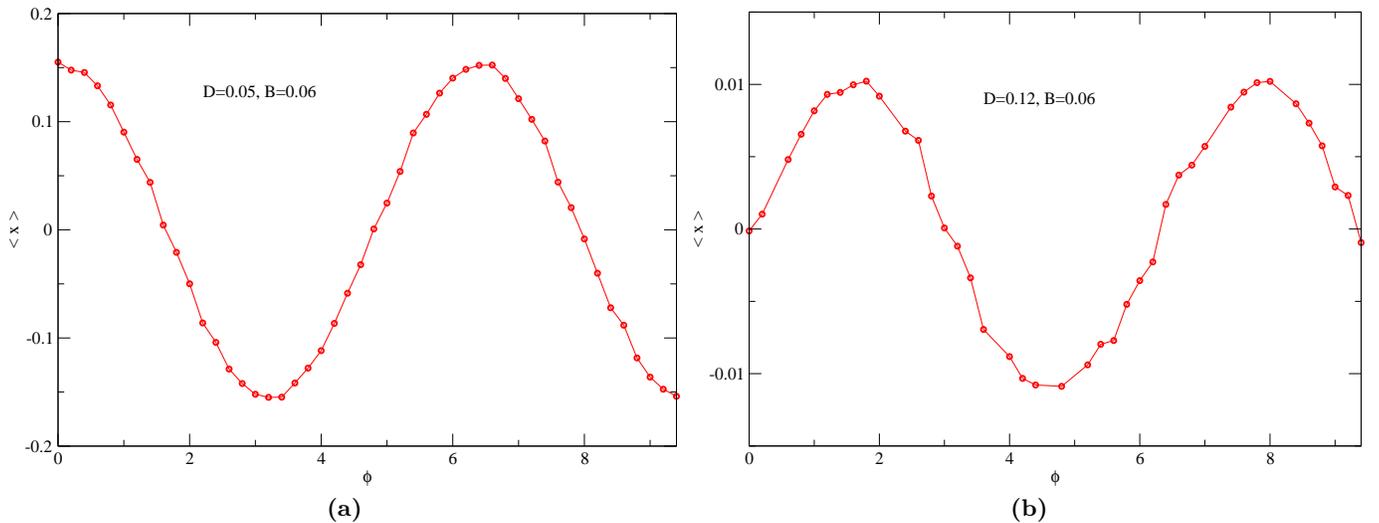

\centering
\begin{tabular}{cc}
\epsfig{file=fig10a.eps, width=0.5\lw, clip=}&
\epsfig{file=fig10b.eps, width=0.5\lw,height=0.355\lw, clip=}\\
{\bf (a)} & {\bf (b)}
\end{tabular}
\caption{Average position $\la x\ra$ as a function of phase $\phi$ for two different values of $D$. In (a) $D=0.05$, and in (b) $D=0.12$. Other fixed parameters are: $B=0.06$, $A=0.1$, and $\omega=0.1$}
\label{fig:xbar}
\end{figure}
\begin{figure}
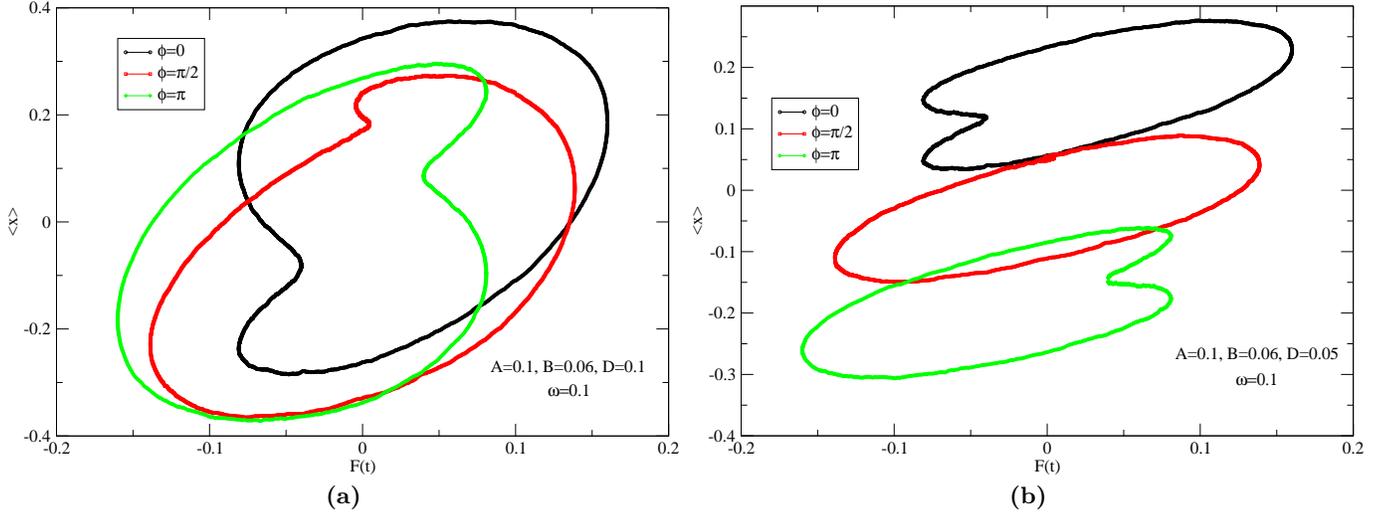

\centering
\begin{tabular}{cc}
\epsfig{file=fig11a.eps,angle=0,width=0.5\linewidth,clip=} &
\epsfig{file=fig11b.eps,angle=0,width=0.5\linewidth,height=0.35\lw,clip=} \\
{\bf (a)} & {\bf (b)} 
\end{tabular}
\caption{(a) Hysteresis loops for different values of  $\phi$ at $D=0.1$, (b) Hysteresis loops for different $\phi$ at $D=0.05$, with other parameters $A=0.1, ~B=0.06$, and $\omega=0.1$ }
\label{fig:hystphi}
\end{figure}
\begin{figure}
\centering
\begin{tabular}{cc}
\epsfig{file=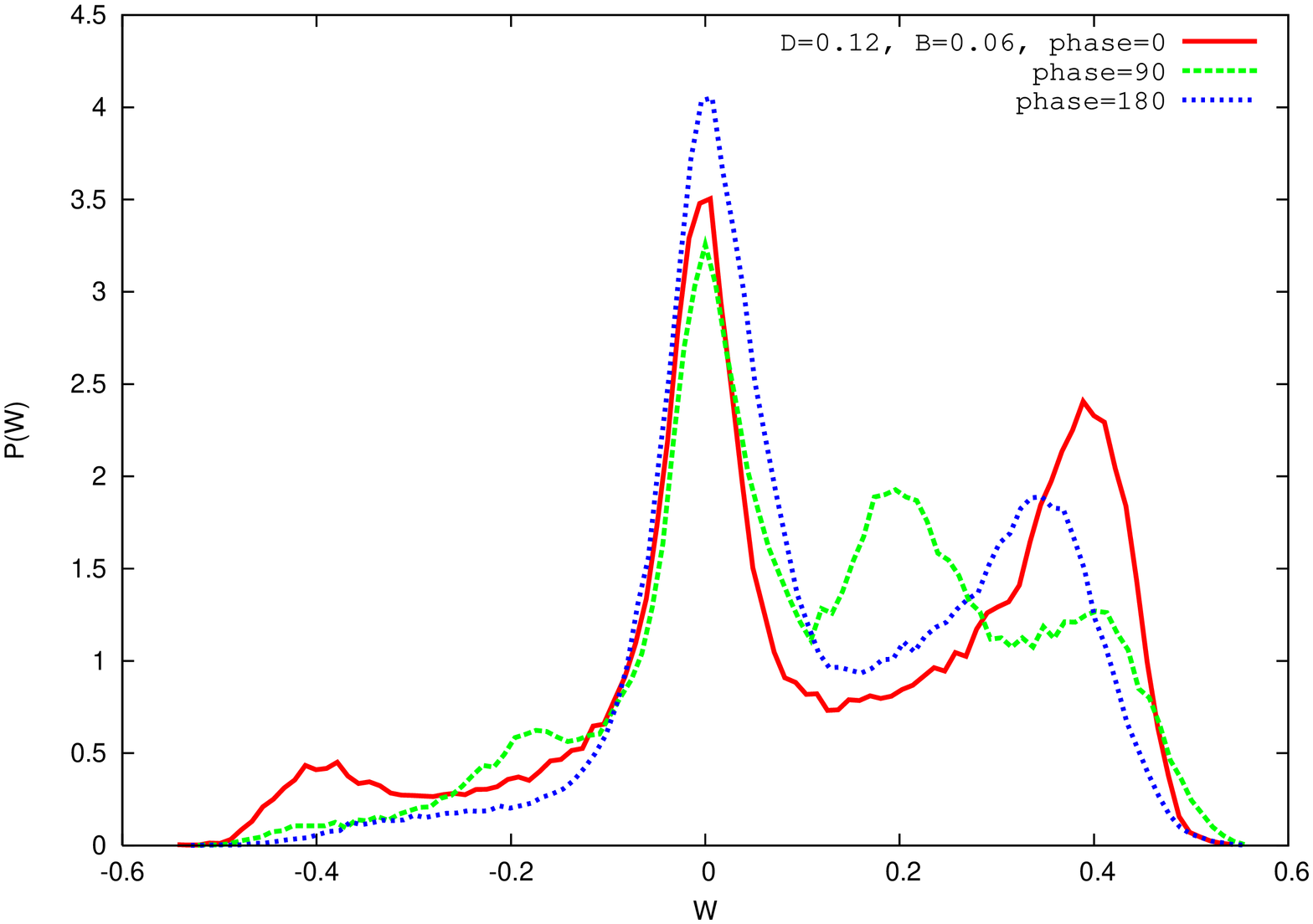,angle=-90,width=0.5\lw,clip=}&
\epsfig{file=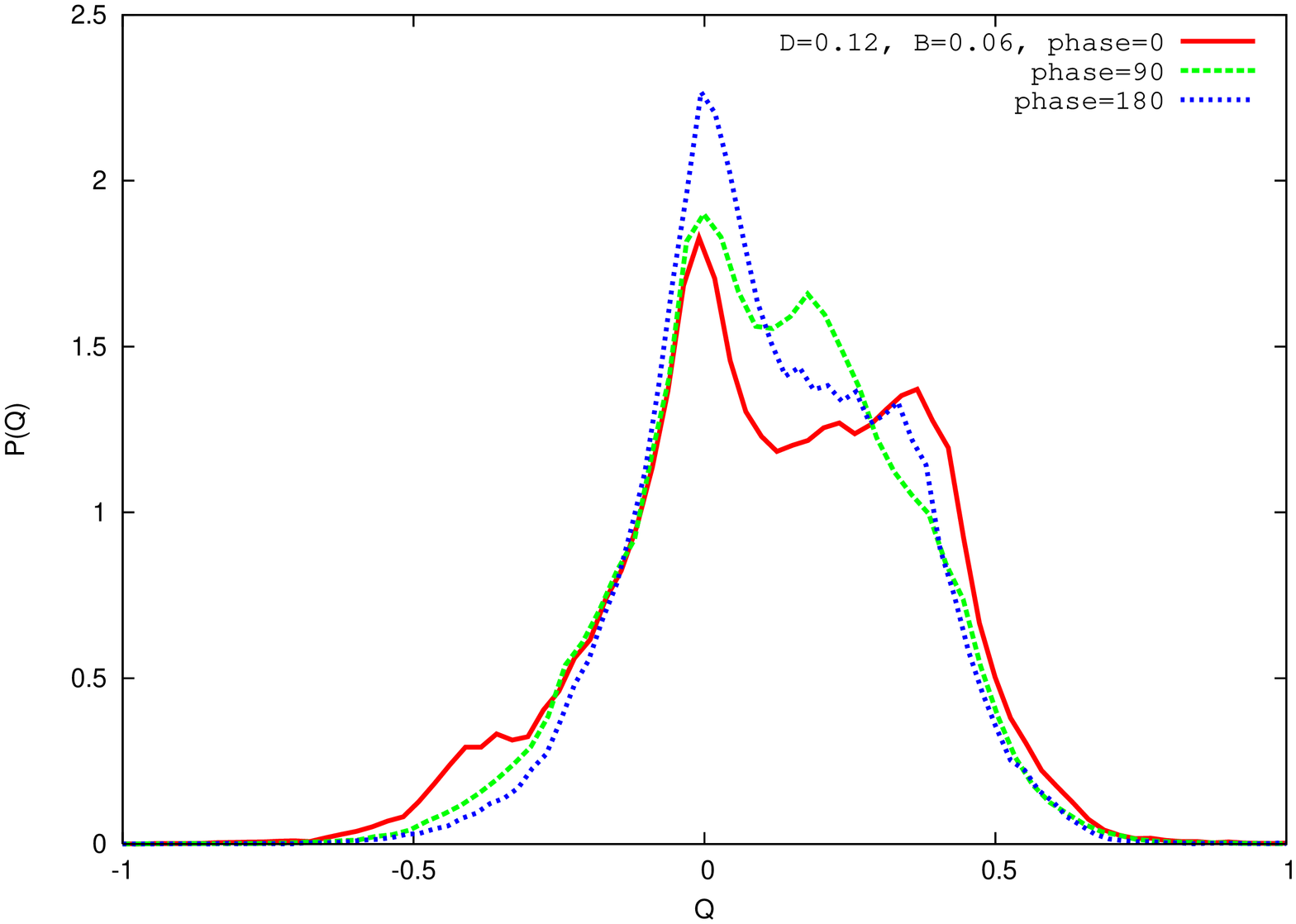,angle=-90,width=0.5\lw,clip=}\\
{\bf(a)} & {\bf(b)} \\
\end{tabular}
\caption{Figures (a) and (b) show the distributions $P(W_p)$ and $P(Q_p)$  respectively for three different values of $\phi$ as mentioned on the graphs. Here  $D=0.12$ and $B=0.06$.}
\label{WQdists}
\end{figure}
We now analyze the role of phase difference($\phi$) between driving fields on pumping and energetics of the system.  In figure \ref{fig:WDphi} (a), we have plotted $\la W_p\ra$ as a function of noise strength $D$ for various values of $\phi$. Other physical parameters are held fixed as mentioned in the figure captions. In figure \ref{fig:WDphi} (b), we have plotted relative variance of $ W_p$ as a function of $D$. It is interesting to note that $\la W_p\ra$ is insensitive to $\phi$, even though the relative variance depends on $\phi$. This is a rather surprising result, given the fact that different values of phase $\phi$ lead to different degrees of localization of the particle in one of the wells. We have characterized this dynamic localization of particles by average position $\la x\ra$ of the particle in the double well potential which in fact can be large depending on $D$ and $\phi$. This is shown in figure \ref{fig:xbar} where we have plotted $\la x\ra$ as a function of $\phi$ for two different values of noise strength $D$. One can readily see that $\la x\ra$ is periodic in $\phi$ as expected. 
The insensitivity of $\la W_p\ra$ on phase gets reflected in the hysteresis loop areas as shown in figures \ref{fig:hystphi} (a)($D=0.1$) and (b)($D=0.05$) for different values of $\phi$ and fixed value of $B$($B=0.06$). We notice that the areas of the hysteresis loops remain same for different $\phi$. However, their shapes are asymmetric and qualitatively different for different $\phi$(i.e., sensitive dependence on phase $\phi$). Due to the different degree of localization or pumping, loops are shifted in $\la x\ra-F$ plane. 
The sensitivity of full probability distribution on the phase difference can be seen from  figures  \ref{WQdists}. In these figures we have plotted $P(W_p)$ and $P(Q_p)$ for different values of $\phi$ as indicated. Note that the distributions exhibit qualitative differences for different $\phi$. We have also verified separately that for different values of rocking amplitudes, as long as we are in subthreshold regime, average input energy is not very sensitive to $\phi$ as opposed to full probability distribution and hysteresis loops. By tuning $\phi$, one can achieve different degrees of particle confinement and can control the fluctuations in heat and work.
\subsection{Energy fluctuations and SSFT}

\label{sec:3F}

\begin{figure}
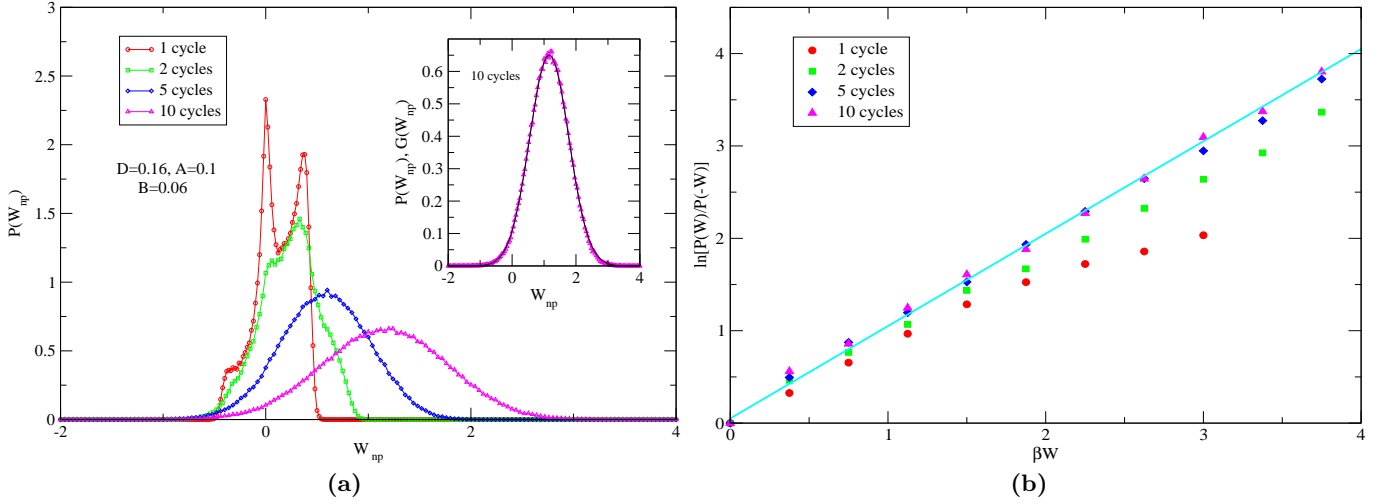

\centering
\begin{tabular}{cc}
\epsfig{file=fig13a.eps,width=0.5\lw,clip=} &
\epsfig{file=fig13b.eps,angle=0,width=0.5\linewidth,height=0.34\lw,clip=}\\ 
{\bf (a)} & {\bf (b)}
\end{tabular}
\caption{(a) The evolution of $P(W_p)$ over different periods. In the inset $P(W_{10p})$ is plotted together with its Gaussian fit $G(W)$. (b) The plot of symmetry function $(\ln{\frac{P(W)}{P(-W)}})$ versus $\beta W$ for various values of periods. The parameters used are: $D = 0.16, \omega = 0.1,  B= 0.06, A = 0.1, \phi = 0.$ The solid line is the best fit for  symmetry function calculated for 10 cycles.}
\label{diffperW}
\end{figure}
\begin{figure}
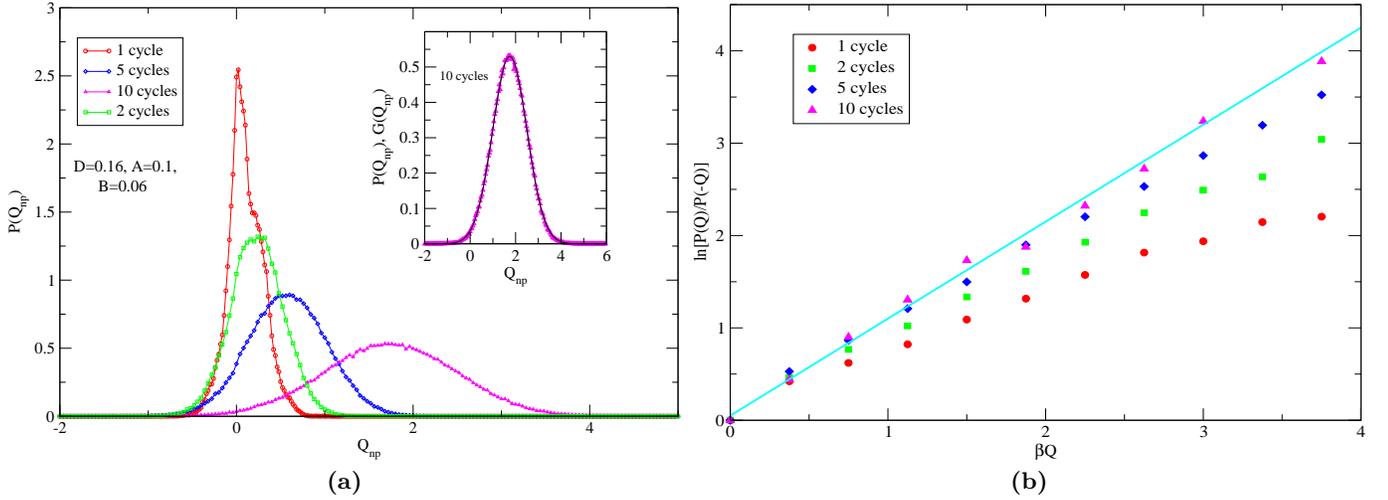

\centering
\begin{tabular}{cc}
\epsfig{file=fig14a.eps,width=0.5\lw,clip=} &
\epsfig{file=fig14b.eps,width=0.5\linewidth,height=0.34\lw,clip=}\\
{\bf (a)} & {\bf (b)}
\end{tabular}
\caption{(a) The evolution of $P(Q_p)$ over different periods. In the inset $P(Q_{10p})$ is plotted together with its Gaussian fit $G(Q)$. (b) The plot of symmetry function $(\ln{\frac{P(Q)}{P(-Q)}})$ versus $\beta Q$ for various values of periods.  The solid line is the best fit for  symmetry function calculated for 10 cycles. The parameters used are same as in figure 13.}
\label{diffperQ}
\end{figure}
Finally we discuss the validity of SSFT in the present case of nonequilibrium time periodic steady state. SSFT implies the probability distribution of physical quantity $A$ to  satisfy relation $p(A)/p(-A) = e^{\beta A}$, where $\beta$ is the inverse temperature of the bath and $A$ is the work done on the system or the heat released to the bath over a long time of observation. For  linear driven systems SSFT for work is indeed satisfied even for work done over a single period in the time asymptotic regime \cite{mam1}. However, for nonlinear systems it has been observed experimentally and theoretically that SSFT is satisfied if one considers work done over a large numbers of cycles \cite{saik,mam,cili}. In regard to heat, SSFT is known to be valid for $Q<\langle Q\rangle$. Since $\langle Q\rangle$ increases with the number of periods or measured time interval in the limit of large n $(n\rightarrow \infty)$, $\langle Q\rangle\rightarrow \infty$ and hence the conventional SSFT is valid over an entire range of Q \cite{joub}. It may be noted that there exists an alternative relation for heat fluctuation,
namely extended heat fluctuation theorem \cite{zon02,zon04}. 
In figure \ref{diffperW} we have plotted probability distribution $P(W_{np})$of work $W_{np}$ integrated over different number (n) of periods. $P(W_{np})$ for a single period exhibits double peak structure. As we increase the number of periods the probability distribution shifts towards right as the mean value of work scales linearly with n. Fine structure in probability distributions get smeared out progressively and distribution tends towards a Gaussian. In the inset of figure \ref{diffperW} (a) we have shown the Gaussian fit for the obtained distribution for 10 cycles. From this fit we obtain $\la W_{10p}\ra = 1.16$ and variance $\sigma^2 \equiv \la W^2\ra - \la W\ra^2 = 0.37$. From this we can obtain dissipation ratio $R_{diss}=\frac{\la W^2\ra-\la W\ra^2}{2\la W\ra/\beta} \simeq 1$, i.e., the variance equals $\frac{2}{\beta}\la W\ra$ which is the required condition to satisfy SSFT when observed distribution is Gaussian \cite{saik,mam1,mspramana}. The validity of SSFT for work is also obviously seen from figure \ref{diffperW} (b) where we have plotted the symmetry function $(\ln{\frac{P(W)}{P(-W)}})$ versus $\beta W$ for work evaluated over different cycles as indicated in the figure. As we increase the number of periods from 1 to 10 the slope of symmetry function approaches 1. Hence our data suggests that the SSFT is satisfied even for finite number of periods. The number of periods above which SSFT is valid depends sensitively on the parameters in the problem.   
As already noted heat fluctuations over a cycle are large compared to work fluctuations. The heat fluctuations get an additional contribution from the internal energy (eq. (\ref{firstlaw})). Even for a linear problem of harmonic oscillator, heat distribution measured over a single period does not satisfy SSFT as opposed to work fluctuations. The contribution from internal energy is supposed to dominate at very large values of Q, making the distribution $P(Q)$ exponential in the large Q limit \cite{zon02,zon04}. However, it may be noted that the distribution of the change in internal energy does not change with number of periods. Heat being an extensive quantity in time, distribution changes as we change the number of periods as shown in figure \ref{diffperQ}(a) where we have plotted $P(Q_{np})$ for various values of $n$. As anticipated, by increasing $n$ distribution tends towards the a Gaussian (see for n=10 cycles). The Gaussian fit for the $P(Q_{np})$ (inset of figure \ref{diffperQ} (a)) gives the value for the variance as 0.56, and mean as 1.74. Thus dissipation ratio is 0.99, which is closer to unity, satisfying SSFT. In principle, one should be able to observe exponential tails for the distribution $P(Q)$ in the large Q limit \cite{zon04}. However, our simulations will not be able to detect it due to lack of required precision. As mentioned earlier, in the limit $n \to \infty$, conventional SSFT holds for heat distributions \cite{cili}. In figure \ref{diffperQ}(b), we have plotted the symmetry functions $(\ln{\frac{P(Q)}{P(-Q)}})$ as a function of $\beta Q$. The slope of the symmetry function approaches unity as we increase $n$, thereby suggesting the validity of SSFT.
\section{Conclusion}
In conclusion, we have studied the nature of energy fluctuations in a biharmonically driven bistable system. This system is driven simultaneously with two periodic input signals of frequencies $\omega$ and $2\omega$, having phase difference $\phi$ between them. The presence of additional periodic drive induces particle confinement  or localization in a preferred potential well. The degree of confinement analyzed in terms of the averaged value of the particle position $\la x\ra$ depends on the system parameters. We have shown that in spite of confinement, SR signal when quantified via the averaged work per period exhibits enhanced response. This is in sharp contrast to the case when confinement is induced by static tilt, which degrades SR. Surprisingly, the average input energy over a period is not very sensitive to $\phi$ even though variation of $\phi$ leads to significant particle pumping. However, changes in $\phi$ does affect qualitatively the nature of hysteresis loop and distributions/fluctuations of work and heat. We have analyzed the fluctuations in work done, heat dissipated, and internal energy over a large but finite number of periods. Our data suggests that the SSFT for work and heat hold in this system.

\section{Acknowledgment}
One of us (AMJ) thanks DST, India for financial support. We also thank Dr. Mangal C. Mahato for his suggestions throughout the work.




\begin{thebibliography}{100}

\bibitem{Bustamante} C.~Bustamante,~J.~Liphardt  and  F.~Ritort,  Physics
Today {\bf 58}, 45 (2005).

\bibitem{evans02}  D.~J.~Evans  and  D.~J.~Searls,   Adv.Phys.  {\bf 51},
1529 (2002)

\bibitem{harris07} R.~~J.~~Harris and G.~~M.~~Sch\"{u}tz, J. Stat. Mech.,
p07020 (2007).

\bibitem{ritort} F. Ritort, Sem. Poincare {\bf 2}, 63 (2003)
\bibitem{kur07}J. Kurchan, J. Stat. Mech, P07005 (2007).

\bibitem{dje93} D. J. Evans, E. G. D. Cohen and G. P. Morriss, Phys. Rev.
Lett. {\bf 71}, 2401 (1993); {\bf 71}, 3616 (1993) [errata].

\bibitem{dje94} D. J. Evans and D. J. Searles, Phys. Rev. E {\bf 50}, 1645
(1994).

\bibitem{galco95}G. Gallavotti and E. G. D. Cohen, Phys. Rev. Lett. {\bf
74}, 2694 (1995); J. Stat. Phys. {\bf 80}, 931 (1995).

\bibitem{kur98}J. Kurchan, J. Phys. A: Math. Gen. {\bf 31}, 3719 (1998).

\bibitem{lebo99} J. L. Lebowitz and H. Spohn, J. Stat. Phys. {\bf 95}, 333
(1999).

\bibitem{crooks99}G. E. Crooks, Phys. Rev. E {\bf 60}, 2721 (1999).

\bibitem{crooks00} G. E. Crooks, Phys. Rev. E {\bf 61}, 2361 (2000).

\bibitem{udo}U. Seifert, Eur. Phys. J. B {\bf64} 423 (2008).

\bibitem{jar} C.~Jarzynski, Phys.Rev. Lett. {\bf 78}, 2690 (1997)
;~~~Phys.~Rev.~ E  {\bf 56}, 5018 (1997).

\bibitem{zon02} R.~van~Zon  and E.~G.~D.~Cohen,  Phys.Rev. E  {\bf 67},
046102 (2002).

\bibitem{zon04} R.~van~Zon  and E.~G.~D.~Cohen,  Phys.Rev. E  {\bf 69},
056121 (2004).

\bibitem{abhi} O.~Narayan and A.~Dhar,   J. Phys.A:Math Gen {\bf 37}, 63
(2004).

\bibitem{cil98}S. Ciliberto and C. Laroche, J. Phys. IV France {\bf 8}, 215
(1998).

\bibitem{wang02} G. M. Wang, E. M. Sevick, E. Mittag, D. J. Searles and D.
J. Evans, Phys. Rev. Lett. {\bf 89}, 050601 (2002).

\bibitem{lip}  J.~Liphardt  et.al,    Science. {\bf 296}, 1833 (2002)
;~~~F.~Douarche   et al,   Europhys. Lett. {\bf 70}, 593 (2005)
;~~~G.~M.~Wang   et al,    Phys. Rev. Lett. {\bf 89} (2002).

\bibitem{trep}  E.~M.~Trepangnier  et al,  Proc. Natt. Acad. Sci. {\bf 101},
15038 (2004).

\bibitem{dou}F.~Douarche, S.~Joubaud, N.B.~Garnier, A.~Petrosyan and
S.~Ciliberto Phys. Rev. Lett 97, 140603 (2006);
R.von Zon, S.~Ciliberto and E.G.D.~Cohen, Phys. Rev. Lett
92, 130601 (2004).

\bibitem{joub}S. Joubaud, N. B. Garnier and S. Ciliberto, J. Stat. Mech., P09018 (2007).

\bibitem{saik}S. Saikia, R. Roy and A. M. Jayannavar,
Phys. Lett A {\bf 369}, 367(2007).

\bibitem{mam}M. Sahoo, S. Saikia, M. C. Mahato and A. M.
Jayannavar, Physica A {\bf 387} 6292 (2008).

\bibitem{cili} P. Jop, S. Ciliberto, A. Petrosyan, Europhys.
Lett. {\bf 81}, 50005 (2008).

\bibitem{parisi} R. Benzi, G. Parisi, A. Sutera and A. Vulpiani, Tellus {\bf
34}, 10 (1982).

\bibitem{gam}L.~Gammaitoni, P.~Hanggi, P.~Jung and F.~Marchesoni, Rev.Mod.
Phys.{\bf 70}, 223 (1998).

\bibitem{gamma}L.~Gammaitoni, F.~Marchesoni and S.~Santucci, Phys. Rev. Lett
{\bf 74}, 1052 (1995).

\bibitem{maha97}M.C.~Mahato and A.M.~Jayannavar, Phys. Rev. E {\bf 55}, 6266
(1997).

\bibitem{mpl}M.C.~Mahato and A.M.~Jayannavar, Mod. Phys. Lett. B {\bf 11},
815 (1997).

\bibitem{evs}M.~Evstigneev, P.~Reimann, C. Schmitt and C.~Bechinger, J.Phys. C {\bf 17},
S3795 (2005).

\bibitem{mahato}M.C.~Mahato and A.M.~Jayannavar, Physica A {\bf 248}, 138
(1998).

\bibitem{iwai}T.~Iwai, Physica A {\bf 300}, 350 (2001).

\bibitem{dan}D.~Dan and A.M.~Jayannavar, Physica A {\bf 345}, 404 (2005).

\bibitem{borro04}M. Borromeo and F. Marchesoni, Europhys. Lett. {\bf 68},
783 (2004).

\bibitem{savelev}S. Savelev et al., Europhys. Lett. {\bf 67}, 179 (2004).

\bibitem{borro05}M. Borromeo and F. Marchesoni, Phys. Rev. E {\bf 71},
031105 (2005).

\bibitem{borro-acta}M. Borromeo and F. Marchesoni, Acta Physica Polonica B
{\bf 36}, 1421 (2005).

\bibitem{pla}M. C. Mahato, A. M. Jayannavar, Phys. Lett. A {\bf 209}, 21
(1995).

\bibitem{chi}D. R. Chialvo and M. M. Millonas. Phys. Lett. A {\bf 209}, 26
(1995).


\bibitem{ajdari}A. Ajdari, D. Mukamel, L. Peliti, and J. Prost, J. Phys.
(Paris)-I, {\bf 4}, 1551 (1994).

\bibitem{mam09} M. Sahoo and A. M. Jayannavar, cond-mat/arXiv:0905.3901.
\bibitem{seki} K.~~Sekimoto,~~J.~~Phys.~~Soc.~~Jpn. {\bf 66}, 6335 (1997).

\bibitem{parrando} J. M. R. Parrondo and B. J. De Cisneros, Appl. Phys. A:
Mater. Sci. Process. {\bf 75}, 179 (2002).

\bibitem{risk}H. Risken, {\em The Fokker-Planck Equation: Methods of
Solution and Applications} (Springer-Verlag Berlin, 1989).

\bibitem{mannela} R.~~Mannela ~~ Lecture Notes in Physics,~~vol {\bf 557}
Springer-Verlag,Berlin, 353 (2000).

\bibitem{mam1}A. M. Jayannavar and M. Sahoo, Phys. Rev. E {\bf 75},
032102 (2007).

\bibitem{mspramana}M. Sahoo and A. M. Jayannavar, Pramana -J. Phys. {\bf 70} 201(2008).

\end{thebibliography}
\end{document}